# A BEACONING APPROACH WITH KEY EXCHANGE IN VEHICULAR AD HOC NETWORKS

Mohammed ERRITALI [1], Oussama Mohamed Reda[1], Bouabid El Ouahidi [1]

[1]Mohamed V University – Faculty of Sciences Rabat,
Data mining and networks laboratory
Department of Computer Science

mederritali@yahoo.fr
oussama.reda@fsr.ac.ma
ouahidi@fsr.ac.ma

## ABSTRACT:

*Vehicular Ad-Hoc Networks (VANETs) are special forms of Mobile Ad-Hoc Networks (MANETs) that allows vehicles to communicate together in the absence of fixed infrastructure.In this type of network beaconing is the means used to discover the nodes in its eighborhood.For routing protocol successful delivery of beacons containing speed, direction and position of a car is extremely important.Otherwise, routing information should not be modified/manipulated during transmission without detection, in order to ensure the routing information, messages must be signed and provided with a certificate to attest valid network participants. In this work we present a beaconing protocol with key exchange to prepare the generation of a signature to protect the routing information protocol 'Greedy Perimeter Stateless Routing'.*

## 1. INTRODUCTION:

Vehicular Ad-Hoc Network (VANET) is a special form of Mobile Ad-Hoc Networks (MANET) that allows vehicles to communicate together in the absence of fixed infrastructure. This network is formed according to the appearance and movement of vehicles and it's a set of moving objects witch communicate with each other using wireless networks like IEEE 802.11 and Ultra Wide Band (UWB) [1].

Vehicular Ad-Hoc Networks (VANET) can be used to develop applications that improve the road traffic safety or allow access to the Internet for passengers.

In such networks,'' beaconing" is the means used by vehicles to find the nodes in their neighborhoods, this mechanism is provided by periodic exchange of "beaconing" messages containing the speed, direction and position of a car.

Security remains the fundamental problem of these new networks, which are inherently vulnerable to various types of attacks such as unauthorized access, spoofing, message modification, and denial of service (DoS). [2.3]

It is therefore necessary that the routing information should not be modified / manipulated during transmission without detection that is why messages should be signed and accompanied by a certificate to prove the validity of the participants in communication in vehicular ad hoc network. In this paper, we present an approach for discovering neighbors with key exchange to prepare the generation of a digital signature to protect the routing protocol Greedy Perimeter Stateless Routing [6].

## 2. KEY EXCHANGE:

Key distribution is a cryptographic technique for exchange of secret keys whose confidentiality is guaranteed by mathematical laws as fast exponentiation.

Among the uses of fast exponentiation in finite fields, we find the secret key exchange; this exchange is done on a network in considering to use a symmetric cryptosystem as AES.

### 2.1 The key agreement: Diffie-Hellman.

The Diffie-Hellman is a technique for exchange of secret keys that can be used to encrypt a conversation. The protocol Diffie-Hellman based on a function of the form $K = W^x \bmod P$ with P prime and W<P.

This function is very easy to calculate, knowledge of K does not allow to deduce X easily.

The two VANET network users Bayan and Nadine each choose a secret number used as the exponent and proceeds as follows:

1. Nadine chooses a number A that will remain his secret.
2. Bayan chooses a number B that will remain his secret.
3. Nadine and Bayan want to exchange the secret key, which is $S = W^{B.A} \bmod P$, but they do

not yet know, since everyone knows that A or B, but not both.

4. Nadine applies to A the one-way function, α is the result: $\alpha = W^A \bmod P$

5. Bayan applies to B the one-way function, β is the result: $\beta = W^B \bmod P$

6. Nadine sends α to Bayan, and Bayan sends β, these two parameter may be known to the whole world without the secret of Nadine and Bayan is disclosed.

7. Nadine received β and calculates $\beta^A \bmod P$ (that is to say in passing by $(W^B)^A \bmod P$ but

he does not know B): $S = \beta^A \bmod P$

8. Bayan received α and computes $\alpha^B \bmod P$ (that is to say in passing by $(W^A)^B \bmod P$, but

he does not know A): $S = \alpha^B \bmod P$

Bayan and Nadine get to the end of their respective calculations the same Key S.

## 3. BEACONING :

In vehicular ad hoc networks beaconing is one of communication modes designed to announce the presence of vehicles in the neighborhood. This neighborhood is detected by the periodic sending and listening to the beacon packets.

When a node starts a routing process using the protocol ''Greedy Perimeter Stateless Routing''.It sends a beacon packet and continues to send these packets at regular intervals.

We describe in the following sections the Beacon algorithm of the protocol ''Greedy Perimeter Stateless Routing'' [4].

### 3.1 Beacon algorithm

The algorithm allows a beacon node to have the locations of its neighbors. Periodically, each node sends a beacon containing its own identifier and location by using two four-byte floating point values for x and y. If a node doesn't receives a beacon packet from a neighboring node after a certain period of time, the GPSR router assumes the neighbor is gone and will remove it from the table of valid neighbors.

The sequence diagram below illustrates the steps of the beacon algorithm:

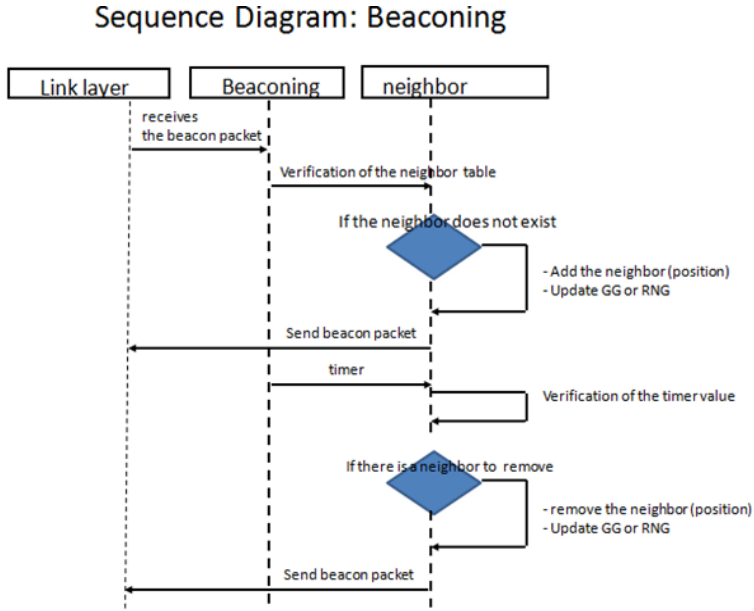

Figure 1. The steps of the algorithm beacon

## 3.2 Detection of neighborhood and key exchange

In this section we present the method for detecting neighbor and exchange a secret key used to generate a symmetric signature that is protection against attacks that target the GPSR routing protocol.

For that a vehicle C detects its neighbor D two messages Suffice: one message for research is broadcasted by C and an acknowledgment is returned by D.

This exchange is illustrated in figure 2:

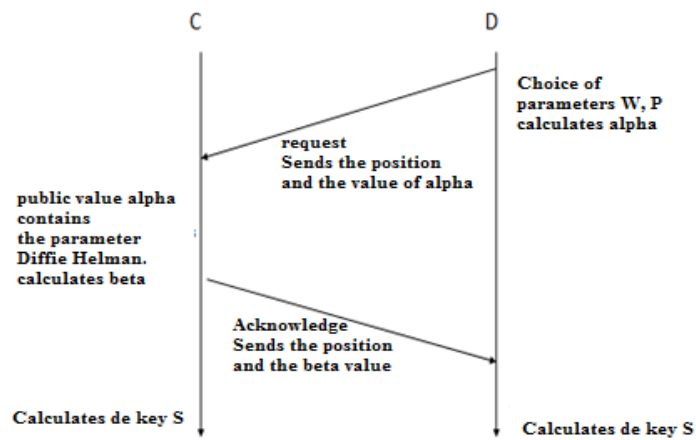

Figure 2. Detection of neighborhood and key exchange

Neighborhood relations are likely to evolve with the movement of vehicles, the neighborhood changes expressed by the appearance of a new neighbor or disappearance of a neighbor.
To see this change the vehicles are required to regularly repeat the process of Neighbor Discovery.

### 3.3 Form of the packet Beacon

In this section we present the shape of the beacon packet used in our experiments:

```
public class Hdr_packet implements java.io.Serializable {

    int identifiant ;
    byte    version;
    byte    type;
    short   packet_len;
    Position  src_pos;
    byte [] publicValue;
}
```

### 4.4 Experimental results:

The various tests are done on a machine with an Intel Pentium (R) 2.13GHz and 2GB of RAM, using JDK 1.7.0.
The following table summarizes the results:

Table 1. Experimental results

| Operation | time in nanosconde |
|---|---|
| Generation parameters and the public value of Diffie-Hellman | 3509800629 |
| Calculate the sender secret key | 49069788 |
| Calculate the receiver secret key | 36127233 |

Our test demonstrates that the operation of neighbor detection and key exchange takes to 4 seconds.

## 4.    CONCLUSION AND  FUTURE WORKS

Security in vehicular ad hoc networks attract more and more attention from research groups, but work in this area and especially in the field of security are still relatively modest. In this work we studied the security problems of routing protocols for vehicular ad hoc networks, then we were interested in the neighbor discovery and secret key exchange for preparing symmetrical signature generation to protect the routing protocol "Greedy Perimeter Stateless Routing".
Neighbor detection is performed by the periodic exchange of beacon message, after finding that this neighborhood is changing with the movement of vehicles; we have suggested how this secret key exchange combined with the message exchange beacons.
As continuity of the work presented, we can deepen our study to improve the emission frequency of beacon messages. Indeed, it is necessary that each node adjusts its transmit frequency on the basis not only of his mobility but also in terms of network density. For this, we will add a parameter to our simulations, the ratio of detection, which will change the transmission frequency according to the density of our vehicular ad hoc network.

# AUTHORS

**Bouabid  El ouahidi**

The Head of computer sciences department and the Network and Data Mining laboratory of the faculty of sciences, Mohamed V University. Obtained a PhD in Computer Sciences from the University of Caen at France in 1992 and a PhD in Computer Sciences from the Mohamed V University at Morocco in 2002. His current interests include developing specification and design techniques for use within Intelligent Network.

**Oussama Mohamed Reda**

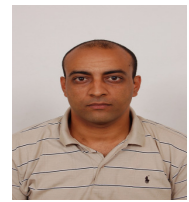

Member of  Network and  Data Mining laboratory and  professor  in the faculty of sciences of  Mohamed V University in Rabat where he has obtained his PhD in open distributed systems in 2009. His current interests include Mobile computing and networking, intelligent transportation and business intelligence systems.

**Mohammed  ERRITALI**

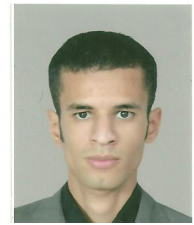

PhD student and member of the Network and Data Mining laboratory of the faculty of sciences, Mohamed V University, Rabat. Obtained a master's degree in business intelligence from the faculty of science and technology, Beni Mellal at Morocco in 2010.  His current interests include developing specification and design techniques for use within Intelligent Network, data mining and cryptography.